\documentclass[mathleft
]{an}
\usepackage{graphicx}
\usepackage{times}
\Pagespan{1}{}
\begin{document}

\Pagespan{99}{}
\Yearpublication{2011}%
\Yearsubmission{2010}%
\Month{99}%
\Volume{999}%
\Issue{99}%

\title{Compact radio emission in Ultraluminous X-ray sources}

\author{M. Mezcua\thanks{
  \email{mmezcua@mpifr.de}\newline Member of the International Max Planck Research School (IMPRS)
	for Astronomy and Astrophysics at the Universities of Bonn and Cologne.}
\and  A.P. Lobanov
}
\titlerunning{Instructions for authors}
\authorrunning{M. Mezcua \& A.P. Lobanov}
\institute{
Max-Planck-Institut f\"ur Radioastronomie,
              Auf dem H\"ugel 69, 53121 Bonn, Germany}

\received{}
\accepted{}
\publonline{later}

\abstract{We present results from our studies of radio emission from
selected Ultraluminous X-ray (ULX) sources, using archival Giant
Metrewave Radio Telescope (GMRT) data and new European VLBI Network
(EVN) observations.  The GMRT data are used to find possible faint
radio emission from ULX sources located in late-type galaxies in the
Chandra Deep Fields. No detections are found at 235, 325 and 610 MHz,
and upper limits on the radio flux densities at these frequencies are
given. The EVN observations target milliarcsecond-scale structures in
three ULXs with known radio counterparts (N4449-X1, N4088-X1, and
N4861-X2). We confirm an earlier identification of the ULX N4449-X1
with a supernova remnant and obtain the most accurate estimates of its
size and age. We detect compact radio emission for the ULX N4088-X1,
which could harbour an intermediate mass black hole (IMBH) of 10$^{5}$
$M_{\odot}$ accreting at a sub-Eddington rate.  We detect a compact
radio component in the ULX N4861-X2, with a brightness temperature $>$
$10^{6}$ K and an indication for possible extended emission. If the
extended structure is confirmed, this ULX could be an HII region with
a diameter of 8.6 pc and surface brightness temperature $\geq10^{5}$
K. The compact radio emission may be coming from a $\sim10^{5}$
M$_{\odot}$ black hole accreting at 0.1$\dot{M}_{Edd}$.}


\maketitle

\section{Introduction}
Several scenarios have been proposed to explain the high luminosities
($L_{X} > 10^{39}$ erg/s) of Ultraluminous X-ray sources (ULXs), but
none of them is able to reveal the physical nature of all ULXs.  If
ULXs are powered by accretion at the Eddington rate, this would
imply accreting compact objects of masses $10^2$-$10^5$
$M_\odot$. Such intermediate compact objects can only be black holes
(Colbert $\&$ Mushotzky \cite{colbert}) and they would be the missing
link between stellar mass black holes and supermassive black holes in
the nuclei of galaxies.  These Intermediate Mass Black Holes (IMBHs) could form from the death of very massive and hot stars or from multiple stellar interactions in dense stellar clusters (Portegies Zwart \cite{portegies}).
It has
also been suggested that ULX objects may harbour secondary nuclear black holes
in post-merger galaxies (Lobanov \cite{lobanov}), with masses in
excess of $10^5$ $M_\odot$ and accreting at sub-Eddington rates. 
Alternatively, ULXs could be neutron stars or stellar mass BHs apparently radiating at super-Eddington luminosities (Begelman \cite{begelman}).

Radio observations of ULXs bear an excellent potential for uncovering the nature of these objects,
by detecting and possibly resolving their compact radio emission, measuring its brightness
temperature and spectral properties, and assessing the physical mechanism for its production.
Few ULXs have been studied in the radio domain (Kaaret et al. \cite{kaaret}; K{\"o}rding et al. \cite{koerding}) and a small sample of ULXs has been cross-identified in the existing radio catalogs (S\'anchez-Sutil et al. \cite{sanchez}).

An increase of the number of radio detections
and subsequent Very Long Baseline Interferometry (VLBI) studies of
detected radio counterparts could potentially help to clarify the
nature of ULX sources.  With this aim, we 1) analyze archive images of
the Chandra Deep Fields taken with the Giant Meterwave Radio Telescope
(GMRT) looking for faint radio counterparts of the ULX sources located
in this fields; and 2) initiate an European VLBI Network (EVN) program
to detect and study milliarcsecond-scale emission in ULX objects with
known radio counterparts.  In Section~2, we present the two samples of
ULX objects studied. The observations and data reduction are explained
in Section~3. The results obtained are shown in Section~4, leading to
a discussion and final summary presented in Section~5.

Throughout this paper we assume a $\Lambda$ cold dark matter (CDM)
cosmology with parameters $H_{0}\ =\ 73\ km\ s^{-1}\ Mpc^{-1}$,
$\Omega_{\Lambda}\ = 0.73$ and $\Omega_{m}\ =\ 0.27$.

\section{Observing targets}
We use archival GMRT data to search for radio counterparts of 24
ULX objects identified in the Chandra Deep Field North
(CDFN), Chandra Deep Field South (CDFS), and Extended CDFS
(Lehmer et al. \cite{lehmer}).  
All these ULXs have luminosities
$L_{X} \ge 10^{39}$ erg/s in the 0.5-2.0 keV band, and are located in
optically bright irregular and late-spiral galaxies. Ten of the 24
X-ray sources appear to be coincident with optical knots of emission,
with optical properties that are consistent with those of giant HII
regions in the local universe, suggesting that these ULX sources trace
distant star formation (Lehmer et al. \cite{lehmer}).

The objects targeted in our EVN observations are selected from a
sample of 11 ULX objects with radio counterparts (S\'anchez-Sutil et
al. \cite{sanchez}) in the VLA FIRST catalog (Becker et
al. \cite{becker}).  We select three ULX which are brighter than 1 mJy
and clearly away from the nuclear region in their respective
host galaxies.  The first target, N4449-X4, is identified as the most
luminous and distant member of the class of oxygen-rich Supernova
Remnants (SNRs) (Blair et al. \cite{Blair83}) and classified as an ULX
source by Liu $\&$ Bregman \cite{liu}. A detailed study of this source
is described in Mezcua $\&$ Lobanov (in prep.). The second
target, N4088-X1, is an ULX located at a distance of 13.0 Mpc in the
asymmetric spiral galaxy NGC 4088. This ULX is located within the
extended emission of a spiral arm and is coincident with a conspicous
maximum of radio emission of 1.87 mJy (at 1.4 GHz) with an offset of
3.62'' to the X-ray peak (S\'anchez-Sutil et al. \cite{sanchez}).  The
ULX has a X-ray luminosity in the 0.3-8.0 keV band of 5.86 x $10^{39}$
erg/s (Liu $\&$ Bregman \cite{liu}, who don't rule out the
possibility of it being an HII region).  The third target, N4861-X2, is
located in the spiral galaxy NGC\, 4861 and has an X-ray
luminosity of 8.4 x $10^{39}$ erg/s (Liu $\&$ Bregman \cite{liu}).
Its radio counterpart is offset from the X-ray position by 1.97''.
This ULX has been suggested to coincide with an HII region powered by
massive early OB type stars (Pakull $\&$ Mirioni \cite{pakull}).

\section{Observations and data reduction}
We use archival GMRT data of the Hubble Deep Field North (overlapping
with our CDFN region of interest) at 235 MHz (experiments 01NIK04 $\&$
11TMA01) observed on 2001 December 17, 2002 January 4 and 2002 January
26, and of the CDFS at 325 $\&$ 610 MHz (experiments 03JAA01 $\&$
11RNA01), observed on 2003 February 13-17 and 2007 February 11-12,
respectively.  The observations were carried out using the standard
GMRT phase calibration mode and spectral line mode.  The data were
analyzed with the NRAO Astronomical Image Processing System (AIPS).
In order to detect very faint sources, we first image the entire
primary beam area and extract all strong point-like objects. Then the
cleaned field of strong sources is subtracted from the visibility
data, and deeper imaging is performed. Finally self-calibration is
applied to increase the dynamic range of the resulting image.  At 610
MHz, 3 pointings of the same field (CDFS) are observed.

The EVN observations (project codes EM072A $\&$ EM072B) have been made
on June 1st $\&$ 2nd 2009, in two separate blocks of 12 hours in
duration, using 9 antennas (Ef, Jb-1, Cm, Wb, On, Mc, Nt, Tr, Sh) at
the wavelength of 18 cm.  In order to detect weak emission from the
ULX objects, we use the phase-referencing technique, calibrating the phases of the target objects with nearby strong and point-like calibrators.
The data reduction was performed in the
standard way using AIPS. Uniform weighting was used in the imaging,
and tapering of the longest baselines was applied for N4088-X1 and
N4861-X2 to improve the detection of extended emission.

\section{Results}
\subsection{ULX sources in the Chandra Deep Fields}
\begin{figure*}
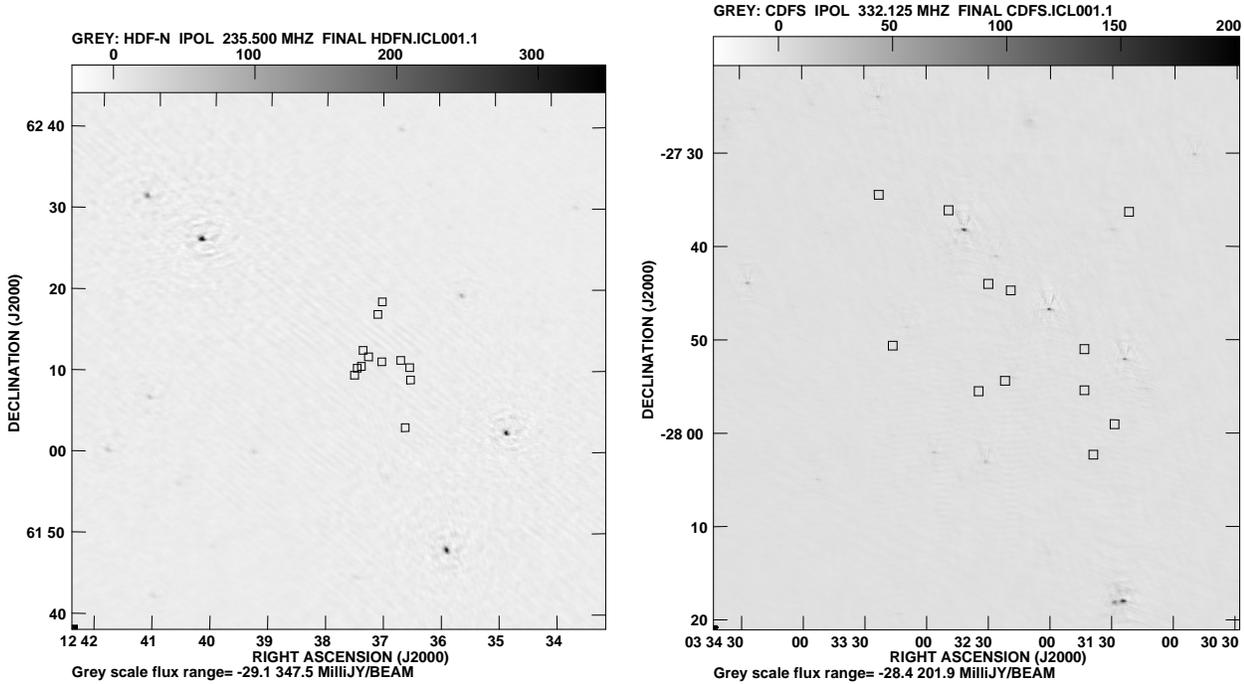

\centering
\includegraphics[width=\columnwidth]{FINAL_HDFN_BOXES.EPS}
\includegraphics[width=\columnwidth]{FINAL_CDFS325_BOXES.EPS}
\caption{GMRT images of the HDFN at 235 MHz (left) and the CDFS at
325 MHz (right). The restoring beam sizes are 20.1 x 16.3 arcsec at 235
MHz and 13.7 x 11.8 arcsec at 325 MHz. The best rms sensitivities
achieved are 1.4, and 0.6 mJy/beam, respectively. The ULXs
positions are marked with squares.}
\label{fig1}
\end{figure*}

We show the final GMRT images obtained at 235 MHz and 325 MHz in Fig.~\ref{fig1}. The
primary beam sizes are 114 arcmin at 235 MHz, 81 arcmin at 325 MHz,
and 43 arcmin at 610 MHz. The respective rms noise in the maps at each
frequency are 1.4, 0.6, and 0.8 mJy/beam. 
No radio counterparts of the ULX sources located in the CDFs are detected in a circle radius for each ULX position of 28 arcsec, which is more than 10 times the Chandra positional error circle. Upper limits on their flux densities at each frequency are given in Table 1 (for ULX located in the CDFN) and Table 2 (ULXs in the CDFS), obtained by estimating the local rms at the ULX locations. These upper limits range between 2-4.6 mJy at 235MHz, 1-2.5 mJy at 332 MHz, and 0.5-2 mJy at 610MHz. The position of only three ULX sources fall in the images at 610MHz, thus only 3 upper limits are given at this frequency.

\begin{table}
\label{table1}
\caption{ULX in the CDFN at 235.5MHz.}
\centering
\begin{tabular}{cc}
\hline
\hline 
Name  & $S$ \\
      & [mJy/beam] \\ 
\hline
CXOHDFN	J123631.66+620907.3	&	$< 3.63$ \\
CXOHDFN	J123632.55+621039.5	&	$< 2.76$ \\
CXOHDFN	J123637.18+621135.0	&	$< 2.88$ \\
CXOHDFN	J123641.81+621132.1	&	$< 3.20$ \\
CXOHDFN	J123701.47+621845.9	&	$< 2.35$ \\
CXOHDFN	J123701.99+621122.1	&	$< 3.12$ \\
CXOHDFN	J123706.12+621711.9	&	$< 2.06$ \\
CXOHDFN	J123715.94+621158.3	&	$< 4.58$ \\
CXOHDFN	J123721.60+621246.8	&	$< 3.47$ \\
CXOHDFN	J123723.45+621047.9	&	$< 4.58$ \\
CXOHDFN	J123727.71+621034.3	&	$< 4.21$ \\
CXOHDFN	J123730.60+620943.1	&	$< 2.11$ \\
\hline
\end{tabular}
\end{table}

\begin{table}
\label{table2}
\caption{ULX in the CDFS and ECDFS at 332MHz and 610MHz.}
\centering
\small
\begin{tabular}{ccc}
\hline
\hline 
Name        & $S_\mathrm{332MHz}$ & $S_\mathrm{610MHz}$ \\
            & [mJy/beam] & [mJy/beam] \\
\hline
CXOECDFS J033122.00-273620.1	&	$< 1.17$	&	...	\\
CXOECDFS J033128.84-275904.8	&	$< 1.38$	&	...	\\
CXOECDFS J033139.05-280221.1	&	$< 1.38$	&	$< 1.62$	\\
CXOECDFS J033143.46-275527.8	&	$< 1.45$	&	...	\\
CXOECDFS J033143.48-275103.0	&	$< 1.63$	&	...	\\
CXOCDFS	J033219.10-274445.6	&	$< 1.85$	&	...	\\
CXOCDFS	J033221.91-275427.2	&	$< 2.39$	&	...	\\
CXOCDFS	J033230.01-274404.0	&	$< 1.63$	&	...	\\
CXOCDFS	J033234.73-275533.8	&	$< 1.28$	&	...	\\
CXOECDFS J033249.26-273610.6	&	$< 2.98$	&	...	\\
CXOECDFS J033316.29-275040.7	&	$< 0.99$	&	$< 0.66$	\\
CXOECDFS J033322.97-273430.7	&	$< 1.08$	&	$< 0.80$	\\
\hline
\end{tabular}
\normalsize
\end{table}

\subsection{Compact radio emission in ULX sources}
\begin{figure*}
\centering \includegraphics[width=\columnwidth]{N4088_CONTOURS.EPS}
\includegraphics[width=\columnwidth]{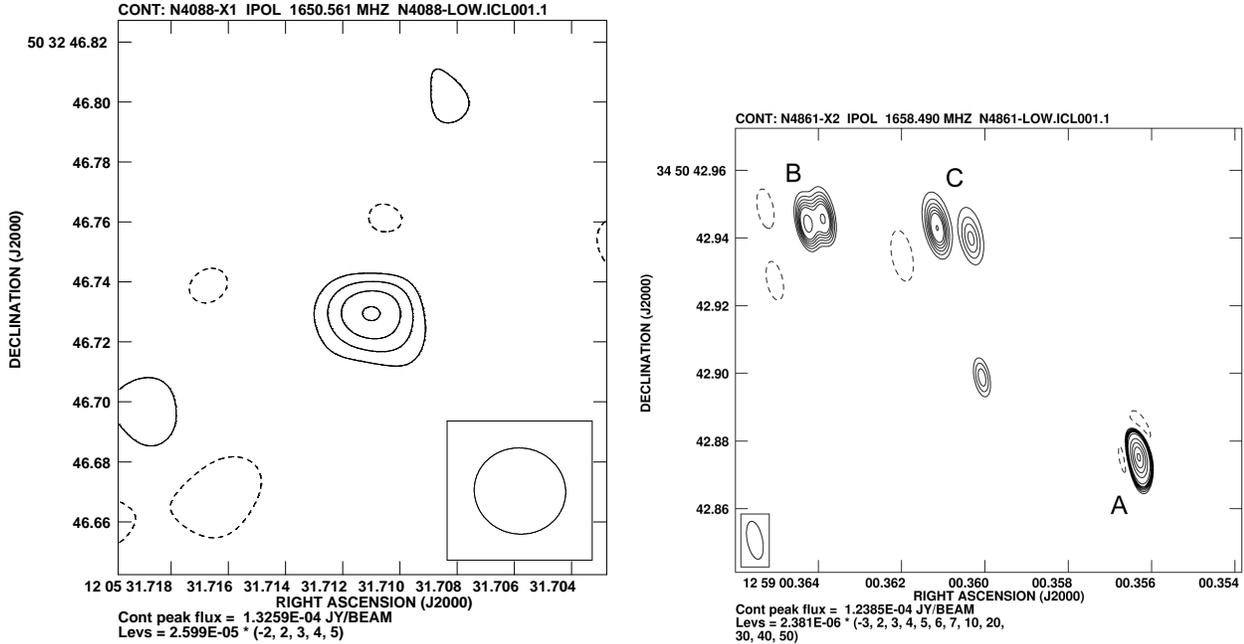}
\caption{1.6 GHz EVN images of the ULX sources N4088-X1 (left) and N4861-X2 (right). The restoring beam sizes are 31 x 29 mas for N4088-X1 and 11 x 5 mas for N4861-X2, with the major axis of the beam oriented along a position angle of 21.99$^{\circ}$ for N4088-X1 and 11.36$^{\circ}$ for N4861-X2. The contours for the left image are (-2, 2, 3, 4, 5) x 26 $\mu$Jy/beam, the rms noise off-source. For the ULX on the right, the contours are (-32, 2, 3, 4, 5, 6, 7, 10, 20, 30, 40, 50) x 3 $\mu$Jy/beam. These radio counterparts are offset from the X-ray peak position by 3.62'' for N4088-X1, and by 1.97'' for N4861-X2. These offsets lie within the X-ray positional error. The compact component of N4861-X2 is indicated with an A. Possible extended emission might be detected in regions B and C. }
\label{fig2}
\end{figure*}

In Fig.~\ref{fig2} we show
the final images of N4088-X1 (left) and N4861-X2 (right). The noise
levels achieved are 26$\mu$Jy/beam for N4088-X1 and 3$\mu$Jy/beam for
N4861-X2, and the restoring beams are 31 x 29 mas and 11 x 5 mas,
respectively. 

For N4088-X1, we identify a compact component of flux
density 0.1 mJy at a 5$\sigma$ level.  The component is centered at \\
RA(J2000) = 12$^h$05$^m$31.7110$^s$ $\pm$ 0.0003$^s$, \\
DEC(J2000) = 50$^{\circ}$32'46.729'' $\pm$ 0.002''.\\
For this component, we estimate
a brightness temperature of T$_{B}$ $>$ 7 x 10$^{4}$ K and an upper
limit of 34 x 26 mas for the size.  Adopting a distance of 13.0 Mpc yields, for N4088-X1, an integrated
1.6 GHz radio luminosity of 3.8 x 10$^{34}$ erg/s.

The ULX N4861-X2 (Fig~\ref{fig2}, right) has a compact component A
centered at \\ RA(J2000) = 12$^h$59$^m$00.3563037$^s$ $\pm$
0.0000008$^s$, \\ DEC(J2000) = 34$^{\circ}$50'42.87500'' $\pm$
0.00002''.\\ It has a flux density of $\sim$80$\mu$Jy (for which we
derive a radio luminosity L$_{1.6GHz}$ = 3.3 x 10$^{34}$ erg/s
assuming a distance to the host galaxy of 14.80 Mpc) and a size upper
limit of 9.8 x 3.8 mas, corresponding to a brightness temperature
T$_{B}$ $>$ 1.1 x 10$^{6}$ K. Two additional components (B and C),
with a total flux density of $\sim$70$\mu$Jy are detected, but
cannot be firmly localized with the present data. If this extention
were confirmed, the whole structure (including component A, B $\&$ C)
would have a total flux density of 0.18 mJy, a luminosity of
L$_{1.6GHz}$ = 7.7 x 10$^{34}$ erg/s and diameter D$\sim$120 mas.
This diameter corresponds to 8.6 pc at the distance
of the host galaxy, and it is in agreement with the typical size of
HII regions found in our Galaxy, like G18.2-0.3 (F$\ddot{u}$rst et
al. \cite{furst}), which has a size of 200 mas, a luminosity of
L$_{1.4GHz}$ = 1.1 x 10$^{33}$ erg/s and is formed by several discrete
sources.

\section{Discussion}
\begin{figure}
\centering
\includegraphics*[width=\columnwidth]{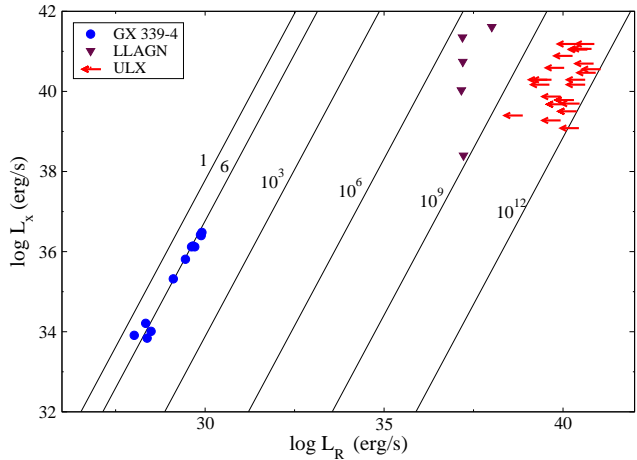}
\caption{Location of the 24 ULXs (red arrows) in the fundamental
plane of sub-Eddington accreting black holes. The parallel lines
correspond to the labeled black hole mass relative to that of the
Sun. We show for comparison the Corbel et al. (\cite{corbel}) data for
the X-ray binary GX 339-4 (filled circles), and the Merloni et
al. (\cite{merloni}) data for the Low Luminosity AGN (LLAGN) NGC 2787,
NGC 3147, NGC 3169, NGC 3226, and NGC 4143 (inverted triangles).}
\label{fig3}
\end{figure}

We use the upper limits obtained from the GMRT data on the radio flux
densities of the ULX objects to locate them in the fundamental plane
of sub-Eddington accreting black holes ({\em cf.}, Corbel et al. \cite{corbel};
Gallo et al. \cite{gallo}; Merloni et al. \cite{merloni}; Falcke et
al. \cite{falcke}) as defined by a correlation between radio core
($L_\mathrm{R}$) and X-ray ($L_\mathrm{X}$) luminosity and black hole
mass, $M_\mathrm{BH}$, $\log L_\mathrm{R}=0.6 \log L_\mathrm{X}+0.78
\log M_\mathrm{BH} + 7.33$.  For our calculations, we assume a radio
spectral index $\alpha_{R}\simeq 0.15$ and a X-ray spectral index
$\alpha_{X}\simeq -0.6$ adopted previously by Falcke et
al. (\cite{falcke}).  The resulting radio and X-ray luminosities of
the ULX objects in our sample are compared in Fig.~\ref{fig3} to the
results of Corbel et al. (\cite{corbel}) and Merloni et
al. (\cite{merloni}). The resulting high upper limits on the BH masses do not
provide strong constraints on the nature of these ULX
objects.

\begin{figure}
\centering
\includegraphics*[width=\columnwidth]{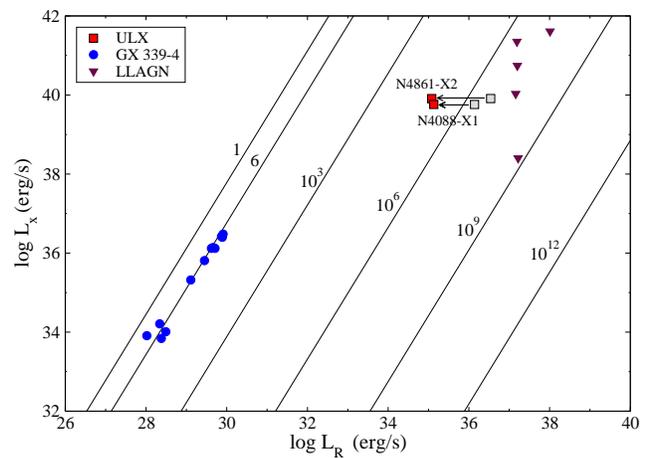}
\caption{Old (S\'anchez-Sutil et al. \cite{sanchez}, grey squares) and new (this work, red squares) location of the ULX sources N4088-X1 and N4861-X2 in the fundamental plane of sub-Eddington accreting black holes.  The parallel lines correspond to the labeled black hole mass relative to that of the Sun. We show for comparison the Corbel et al. (\cite{corbel}) data for the X-ray binary GX 339-4 (filled circles), and the Merloni et al. \cite{merloni} data for the Low Luminosity AGN (LLAGN) NGC 2787, NGC 3147, NGC 3169, NGC 3226, and NGC 4143 (inverted triangles).}
\label{fig4}
\end{figure}

A similar relation is shown in Fig.~\ref{fig4} for the most compact
components in the EVN images of N4088-X1 and N4861-X2. Using our radio
luminosity at 1.6 GHz appropriately scaled to 5 GHz, the X-ray
luminosity scaled to the 2-10 keV band, and assuming a sub-Eddington
accretion regime, we derive a black hole mass of 10$^{5.1}$ M$_\odot$
and of 10$^{4.9}$ M$_\odot$ for N4088-X1 and N4861-X2 (component A),
respectively. These masses are in agreement with the IMBH scenario
for both objects.

Higher sensitivity observations are needed, and observational time has
already been guaranteed, to try to detect and/or confirm possible
extended structure for both N4088-X1 and N4861-X2.

\section{Summary}

Radio observations of ULX sources can help to unveil the nature of
these objects.  Analysis of archival GMRT data of the Chandra Deep
Fields at 235, 325 $\&$ 610 MHz have not yielded any radio
counterparts for these sources but yielded upper limits on their
flux densities.  These ULXs are too weak for deep field radio
observations so higher sensitivity is needed in order to detect any
faint radio emission.  

New EVN observations of three ULXs with known
radio counterparts yielded first milliarcsecond-scale images of all
three objects. 
The EVN observations have confirmed the earlier
 identification of the ULX N4449-X1 with a SNR and obtained the most
 accurate estimates of its size and age (Mezcua $\&$ Lobanov
 in prep.). 
For the two other ULXs studied, N4088-X1 and N4861-X2, the EVN measurements have provided improved
estimates of the compact radio flux, yielding better localizations of
these objects in the L$_{X}$-L$_{radio}$ diagram. The suggested nature
of these objects can be best verified with more sentitive observations
at 5 GHz aimed at both improving the brightness temperature estimates
and obtaining spectral index information. The success of the EVN
observations also calls for expanding this study to more ULX objects.

\acknowledgements

The authors are grateful to M. L{\'o}pez-Corredoira and M.~W. Pakull
for their valuable comments.  M. Mezcua was supported for this
research through a stipend from the International Max Planck Research
School (IMPRS) for Radio and Infrared Astronomy at the Universities of
Bonn and Cologne.



\begin{thebibliography}{}
\bibitem[2002]{begelman} Begelman, M.~C.\ 2002, \apjl, 568, L97 

\bibitem[1995]{becker} Becker, R.~H., White, R.~L., \& Helfand, D.~J.\ 1995, \apj, 450, 559 

\bibitem[1983]{Blair83} Blair, W.~P., Kirshner, R.~P., \& Winkler, P.~F., Jr.\ 1983, \apj, 272, 84 

\bibitem[1999]{colbert} Colbert, E.~J.~M., \& Mushotzky, R.~F.\ 1999, \apj, 519, 89 

\bibitem[2003]{corbel} Corbel, S., Nowak, M.~A., Fender, R.~P., Tzioumis, A.~K., \& Markoff, S.\ 2003, A\&A, 400, 1007 

\bibitem[2004]{falcke} Falcke, H., K{\"o}rding, E., \& Markoff, S.\ 2004, A\&A, 414, 895 

\bibitem[1987]{furst} Fuerst, E., Reich, W., Reich, P., Handa, T., \& Sofue, Y.\ 1987, A\&AS, 69, 403 

\bibitem[2003]{gallo} Gallo, E., Fender, R.~P., \& Pooley, G.~G.\ 2003, \mnras, 344, 60 

\bibitem[2003]{kaaret} Kaaret, P., Corbel, S., Prestwich, A.~H., \& Zezas, A.\ 2003, Science, 299, 365 

\bibitem[2005]{koerding} K{\"o}rding, E., Colbert, E., \& Falcke, H.\ 2005, A\&A, 436, 427 

\bibitem[2006]{lehmer} Lehmer, B.~D., Brandt, W.~N., Hornschemeier, A.~E., Alexander, D.~M., Bauer, F.~E., Koekemoer, A.~M., Schneider, D.~P., \& Steffen, A.~T.\ 2006, \aj, 131, 2394 

\bibitem[2005]{liu} Liu, J.F., \& Bregman, J.~N.\ 2005, \apjs, 157, 59 

\bibitem[2007]{lobanov2007} Lobanov, A.\ 2007, From Planets to Dark Energy: the Modern Radio Universe

\bibitem[2008]{lobanov} Lobanov, A.~P.\ 2008, Memorie della Societa Astronomica Italiana, 79, 1306 

\bibitem[2003]{merloni} Merloni, A., Heinz, S., \& di Matteo, T.\ 2003, \mnras, 345, 1057 


\bibitem[2002]{pakull} Pakull, M.~W., \& Mirioni, L.\ 2002, arXiv:astro-ph/0202488 

\bibitem[2003]{portegies} Portegies Zwart, S.~F.\ 2003, Astrophysical Supercomputing using Particle Simulations, 208, 145 

\bibitem[2006]{sanchez} S{\'a}nchez-Sutil, J.~R., Mu{\~n}oz-Arjonilla, A.~J., Mart{\'{\i}}, J., Garrido, J.~L., P{\'e}rez-Ram{\'{\i}}rez, D., \& Luque-Escamilla, P.\ 2006, A\&A, 452, 739 

\end{thebibliography}
\end{document}